\documentclass[12pt,showpacs,twocolumn, preprintnumbers,aps,pre, reprint,superscriptaddress,nofootinbib]{revtex4-1}

\usepackage{graphicx}
\usepackage{subfigure}
\usepackage[labelfont={footnotesize,bf},textfont={footnotesize}]{caption}
\usepackage{color}
\usepackage{amsmath,amssymb,mathtools}
\usepackage{epstopdf}
\usepackage[linktocpage,colorlinks=true,linkcolor=blue,citecolor=blue, urlcolor = black, breaklinks=true]{hyperref}
\usepackage{url}
\usepackage{verbatim}
\usepackage{breakcites}
\usepackage[version=3]{mhchem}

\newcommand{\be}{\begin{equation}}
\newcommand{\ee}{\end{equation}}

\makeatother

\begin{document}

\preprint{}




\title{Eddy memory as an explanation of intra-seasonal periodic behavior in baroclinic eddies}

\author{Woosok Moon}
\affiliation{Department of Mathematics, Stockholm University 106 91 Stockholm, Sweden}
\affiliation{Nordita, Royal Institute of Technology and Stockholm University, SE-10691 Stockholm, Sweden}
\email[]{woosok.moon@gmail.com}

\author{Georgy E. Manucharyan}
\affiliation{School of Oceanography, University of Washington, Seattle, WA, USA }

\author{Henk A. Dijkstra}
\affiliation{Institute for Marine and Atmospheric Research Utrecht, Utrecht University, Utrecht, Netherlands}

\date{\today}

\begin{abstract}
The baroclinic annular mode (BAM) is a leading-order mode of the eddy-kinetic energy in the Southern Hemisphere exhibiting 
oscillatory behavior at intra-seasonal time scales. The  oscillation mechanism has been linked to transient 
eddy-mean flow interactions that remain poorly understood. Here we demonstrate that the finite memory effect in 
eddy-heat flux dependence on the large-scale flow can explain the origin of the BAM's oscillatory behavior. 
We represent the eddy memory effect by a delayed integral kernel that leads to a generalized Langevin equation 
for the planetary-scale heat equation. Using a mathematical  framework for the interactions between planetary and synoptic scale motions, 
we derive a reduced dynamical model of the BAM  -- a stochastically-forced oscillator with a period proportional to the geometric mean 
between the eddy-memory time scale and the diffusive eddy equilibration timescale. 
Our model provides a formal justification for the previously proposed phenomenological model of the BAM 
and could be used to explicitly diagnose the memory kernel and improve our understanding of transient eddy-mean flow interactions in the atmosphere.

\end{abstract}


\maketitle


\section{Introduction} 
Large-scale atmospheric dynamics at mid-latitudes is significantly influenced 
by the baroclinic wave life cycle initiated by shear instability \citep{pierrehumbert1995}.
Constrained by geostrophic and hydrostatic balance, a meridional temperature 
gradient imposed by the heat imbalance between tropics 
and high latitudes is proportional to the vertical shear of the mid-latitude jet, 
which acts as a key parameter controlling the growth rate of 
baroclinic waves \citep{charney1947, eady1949, phillips195}. 
During the baroclinic growth of synoptic waves, 
poleward heat flux increases and
the waves propagate upward, after which the waves break 
near critical latitudes while propagating equatorward. 
The whole life cycle of unstable baroclinic waves could be 
represented by energy exchange between zonal mean zonal wind
and synoptic eddies \citep{lorenz1955,simmons1978, edmon1980, moon2009}. 
This cycle is reflected in daily weather at mid-latitudes 
and understood as an engine 
to transfer residual 
energy from the tropics to high latitudes. It determines many aspects 
of large-scale circulations in extra-tropical areas \citep{held1985}.  

The life cycle of baroclinic waves is an essential part of variability 
in mid-latitudes with its time-scale being around 3-4 days.
However, the low-frequency intra-seasonal variability with a time scale longer than weather but shorter
than seasonal \citep{branstator1992, hartmann1974} is not entirely explained 
by the traditional baroclinic wave life cycle \citep{pratt1977, barnston1987}. 
For example, the North Atlantic Oscillation (NAO) \citep{hurrell2003}, 
which represents the fluctuation of the difference of atmospheric pressure
between the Icelandic Low and the Azores High, contains the intra-seasonal time-scale of around 10-days. 
Similarly,  the Arctic Oscillation (AO) \citep{thompson1998}
or the Southern Annular Mode (SAM) \citep{limpasuvan1999}, representing the zonally symmetric seesaw 
between sea level pressures in polar and temperate
latitudes in Northern and Southern Hemisphere, respectively, also vary on  intra-seasonal timescales. 

The variability of the mid-latitude westeries is commonly defined using the zonal index \citep{namias1950} 
that was originally defined by the pressure difference between 35N and 55N, 
but there are many variants \citep{robinson1996, li2003}. The zonal index, 
especially in the Southern Hemisphere,  shows strong intraseasonal 
and interannual variability based on several observational data \citep{hartmann1998}. On intra-seasonal 
time-scales, it is hard to identify a major external forcing and hence the mid-latitude jet variability 
could be controlled by internal dynamics  \citep{branstator1992}. 
Indeed, the low-frequency variability of the zonal index has been identified  
in simple numerical models with zonally-symmetric thermal forcing \citep{robinson1991}, 
implying that variability can be  caused by the interaction between synoptic eddies and zonal mean field. 

The eddy-mean flow interaction is a highly nonlinear process involving turbulent mixing of synoptic waves 
leading to energy transfer among different scales that are difficult to accurately parameterize. 
However, phenomenological models of eddy-mean flow interactions exhibiting oscillatory behavior have been put forward. 
\citet{thompson2014} introduced a stochastic model to explain the quasi-oscillatory behavior of the poleward heat flux in the Southern Hemisphere, 
also referred to as the Southern Hemisphere baroclinic annular mode (BAM) introuduced
by \cite{thompson2014a}. 
Specifically, they suggest 
a two-dimensional stochastic model representing interactions between
the poleward heat flux and the meridional temperature gradient to capture 
the quasi-oscillatory variability with a dominant time-period around 25 days. 
In this model, the increase of the poleward heat flux is proportional 
to the growth rate of eddies generated by baroclinic instability, 
which is proportional to the meridional temperature gradient. 
At the same time, the time evolution of the anomalous meridional temperature gradient 
is controlled by the poleward heat flux with a damping. 
This phenomenology results in a two-dimensional 
stochastic dynamical system that contains oscillatory solutions 
depending on the choice of parameters. 
With this idealized description of the variability, \citet{thompson2014}  suggest that the quasi-oscillatory  low-frequency variability 
can be caused by nonlinear eddy-mean flow interaction. Even though the model reflects the basic 
characteristics of the baroclinic wave life cycle, its direct connection to the primitive equations is not described, which obscures
the physical interpretation of model parameters.

The BAM is a companion index of the Southern Annular mode (SAM) in the Southern Hemisphere representing the characteristics of
large-scale low frequencies in the atmosphere. 
The SAM is defined by zonal mean kinetic energy describing the variability of zonal mean wind mainly influenced by the momentum flux.
The BAM is constructed by eddy kinetic energy mainly controlled by the meridional heat flux in lower levels \citep{thompson2014a}. 
The two processes, meridional heat flux in lower levels and momentum flux in upper ones, 
show different characteristics in their variabilities \citep{boljka2018a,boljka2018b,wang2015,wang2016,pfeffer1992}.
The SAM is well approximated by an autoregressive model of order 1 (AR(1) process), 
having a red noise spectrum, but the BAM shows a distinct peak around 25 days in its power spectrum.
The stochastic oscillatory behavior shown in the BAM is worthwhile to be investigated in detail
due to the expectation that it leads to significant progress in predictability 
in mid-latitudes on sub-seasonal time-scales. Furthermore, \cite{thompson2015} argue
that the BAM in the Northern Hemisphere shows a dominant peak located at around 20 days
in its power spectrum.
This implies that the oscillatory behavior in meridional heat flux or eddy kinetic energy is an intrinsic feature
of eddy-mean flow interactions in large-scale atmospheric dynamics.
Therefore, the major question is how the meridional heat flux in lower levels 
can induce the oscillatory behavior in the interaction with mean field.

Generation of low-frequency oscillations in a rotating fluid 
does not seem to be limited to the large-scale atmosphere as similar 
internal oscillations were also found in large-scale ocean currents. 
An eddy-resolving ocean model simulation 
with a repeated annual cycle forcing reveals an intrinsic mode  
in the Southern Ocean with a period of 40-50 year \citep{le2016, van2017}. 
An idealized model of the surface-stress-driven Beaufort Gyre in the Arctic Ocean 
with mesoscale eddies as a key equilibration process \citep{manucharyan2016} generates
an oscillation with a period of 50 years \citep{manucharyan2017}. 
To explain the oscillation \citet{manucharyan2017}  introduce
eddy memory into the relation between the eddy buoyancy flux and the mean buoyancy gradient, 
which leads to a  modification of the commonly used Gent-McWilliams (GM) parameterization \citep{gent1990}.  
The inclusion of eddy memory leads to a stochastic oscillation if the ratio between the memory timescale 
and the diffusive equilibration timescale reaches a certain threshold \citep{manucharyan2017}. 

Generally, according to the Mori-Zwanzig formalism \citep{zwanzig1961}, 
the dynamic interactions between slow-evolving modes and fast ones
are reflected as a delayed integral on the time evolution 
of the slowly-varying modes. The memory effect, 
normally represented by a delayed term or integral, 
is not new in climate science. One of the simplest 
models for the El-Nin\~{o}
and Southern Oscillation (ENSO) is a delayed ordinary 
differential equation \citep{suarez1988}. Even with such one-dimensional model, 
it was found that the delayed model contains chaotic 
dynamics under seasonal forcing\citep{tziperman1994}. Recently, 
such delay equation climate models were derived using 
the Mori-Zwanzig formalism \citep{falkena2019}. 
Considering the complex interactions in fluid-dynamical systems, 
the memory effect is an intrinsic characteristic leading 
to various internal variability. 

Our research focuses on the memory effect in the interaction between synoptic eddies and a zonal mean zonal wind and its relation to 
 low frequency modes on timescales longer than the weather. 
More specifically, we propose a formalized explanation of the quasi-oscillatory behavior of the BAM \citep{thompson2014} 
by incorporating the eddy memory  \citep{manucharyan2017} into multi-scale equations of atmospheric dynamics \citep{moon2020}. 
We outline the multi-scale atmospheric model in  Section \ref{sec:plenatary motion} and we introduce the eddy memory effect 
onto the planetary -scale equations in Section 3., where we formally derive the reduced model for the zonal mean flow, 
the stochastic oscillator. We conclude in Section 4.

\section{\label{sec:plenatary motion} Multiscale model of atmospheric motions}
The basic governing equation for Earth's climate starts 
from the simplest heat flux balance 
between short-wave radiative flux and outgoing longwave radiative flux, 
which gives us a global average temperature as an equilibrium. This is 
possible when we consider the Earth as a point object in the universe 
maintaining a thermal equilibrium state. If we magnify the size of the
Earth from a point to a sphere, we can see that there is an imbalance of 
heat flux from tropics to higher latitudes, 
which requires transferring heat from the tropics to higher latitudes. 
A simple approximation of the poleward heat flux is a turbulent 
diffusion with a constant eddy diffusivity. 
In an even more magnified view of the
tropics and polar areas, the dominant physics shifts from energy flux balance 
to potential vorticity conservation, which 
acts as a theoretical foundation for the generation of mid-latitude weather systems. 
There are unclearly defined timescales between the weather dominated 
by the vorticity dynamics
and the season controlled by external heat flux. 
These timescales lie on around 10 to 30 days, where numerous variability has been detected 
in climate data. 
There should be an intermediate framework lying between 
the energy flux balance and the vorticity dynamics to investigate 
causes of the variability. 

The large-scale atmosphere is governed by the primitive equations 
which are comprised of three momentum balances, the continuity equation 
and the heat budget.
In Cartesian coordinates, the three momentum
equations are
\begin{align}
    &\frac{Du}{Dt}-f v  
    = -\frac{1}{\rho}\frac{\partial P}{\partial x} \nonumber \\
    &\frac{Dv}{Dt}+ f u 
    = -\frac{1}{\rho}\frac{\partial P}{\partial y} \nonumber  \\
    &\frac{Dw}{Dt}
    =-\frac{1}{\rho}\frac{\partial P}{\partial z}-g,
\end{align}
where $P$ and $\rho$ are atmospheric pressure and density, respectively, 
and the velocities in
the zonal, meridional, and vertical directions are 
$u$, $v$, and $w$. The Coriolis parameter $f$
is approximated by $f_0+\beta y$, where $f_0$ is a Coriolis parameter
calculated at a reference latitude $\phi_0$
and $\beta y$ is the latitudinal variation of the Coriolis parameter where $\beta=\partial f/\partial y$.
The material derivative $D/Dt$
is defined as 
$D/Dt \equiv \partial/\partial t+u\partial/\partial x+v\partial/\partial y+w\partial/\partial z$.
Along with the momentum equations,
the continuity equation is
\begin{align}
    \frac{1}{\rho}\frac{D\rho}{Dt}+\frac{\partial u}{\partial x}+\frac{\partial v}{\partial y}
    +\frac{\partial w}{\partial z} = 0.
\end{align}
Finally, to complete the equations, we need the ideal gas 
law  $P=\rho R T$ and thermodynamic equation
\begin{align}
    \frac{D\Theta}{Dt}=\frac{\Theta}{C_P T}\left(\frac{\kappa}{\rho}\nabla^2T+Q\right),
\end{align}
where $\Theta$ is the potential temperature, i.e.,
\begin{align}
    \Theta \equiv T\left(\frac{P_{00}}{P}\right)^{R/C_P},
\end{align}
where $P$ and $T$ represent the pressure and the temperature, respectively, $P_{00}$ is a
reference pressure, typically $1000$mb,
$R$ is the gas constant for air, $C_P$ 
the specific heat at the constant pressure, and $P_{00}$ is a reference pressure.
The full equations are highly nonlinear and complicated
and to explain the BAM, we need to
simplify the equations using an appropriate scaling.

The scale of atmospheric motions is distinguished by the magnitude 
of the Rossby number, $U/fL$, where $U$ is a horizontal
velocity scale, 
and $L$ is a horizontal length scale. When the Rossby number is 
asymptotically small, the geostrophic balance 
becomes a dominant force balance in the horizontal momentum equations.
The large-scale atmospheric flows satisfying the geostrophic and hydrostatic
balances are called geostrophic motions.

There are two kinds of geostrophic motions depending on the horizontal 
length scale \citep{phillips1963}. If the length scale is similar 
to the internal Rossby deformation radius ($L \sim 1000km$), 
the motion is called quasi-geostrophic and is described by the 
conservation of potential vorticity,
\begin{align}
  \frac{{\rm D}_H}{{\rm D} t} \left[\frac{\partial^2 \psi}{\partial x^2} +
      \frac{\partial^2 \psi}{\partial y^2} +
      \frac{1}{\rho_s}\frac{\partial}{\partial z}
      \left(\frac{\rho_s}{S}\frac{\partial \psi}{\partial z}\right) + \beta
      y\right]\ & =0, & 
\end{align} 
where
\begin{align}
 \frac{{\rm D}_H}{{\rm D} t} = \frac{\partial}{\partial t}-\frac{\partial \psi}{\partial y}\frac{\partial}{\partial x}
 +\frac{\partial \psi}{\partial x}\frac{\partial}{\partial y},
\end{align}
$\psi$ is a leading-order pressure field acting as a stream function, $\rho_s$ 
is a mean vertical density profile, $S$ represents 
the average vertical stability, and $\frac{\partial^2 \psi}{\partial x^2} +
      \frac{\partial^2 \psi}{\partial y^2} +
      \frac{1}{\rho_s}\frac{\partial}{\partial z}
      \left(\frac{\rho_s}{S}\frac{\partial \psi}{\partial z}\right) + \beta y$ is the potential vorticity. All the variables are
non-dimensionalized \citep{pedlosky2013}.
On the other hand, if the horizontal length-scale 
is close to the external Rossby deformation radius 
($L_D \sim 3000km$), the governing equations become
\begin{align}
& u_L\ =\ -\frac{\partial P_L}{\partial y} \\ & v_L\ =\ \frac{\partial
    P_L}{\partial x} \\ & \rho_L\ =\ -\frac{1}{\rho_s}\frac{\partial}{\partial
    z}\Big(\rho_s P_L\Big) \\ & \frac{1}{\rho_s}\frac{\partial}{\partial
    z}\Big(\rho_s w_L\Big) - \beta_L v_L\ =\ 0 \\ & \frac{\partial
    \Theta_L}{\partial t} + u_L\frac{\partial \Theta_L}{\partial x} +
  v_L\frac{\partial \Theta_L}{\partial y} + w_L\frac{\partial \Theta_L}{\partial
    z} + \frac{w_L}{\epsilon\Theta_s}\frac{{\rm d}\Theta_s}{{\rm d}z}\ =\ Q_L,
\end{align}
where $P_L$ is the planetary-scale pressure, $u_L$, $v_L$ and $w_L$ are the zonal, 
the meridional and the vertical velocity, respectively, $\Theta_s$ is 
the hemispheric average potential temperature, 
$\Theta_L$ is an anomalous potential temperature, 
$\beta_L$ is the planetary-scale beta effect, and $Q_L$ represents radiative processes.
The thermodynamic forcing $Q_L$ is a residual of local radiative fluxes driving a local temperature change. 
The subscript $L$ implies planetary-scale variables.
Eqs. (7)-(11) are usually referred to as planetary geostrophy, where
the heat equation with
advection is constrained by the geostrophic and hydrostatic 
balances, together with the Sverdrup relation \citep{moon2020}.
Similar results can be found in \cite{dolaptchiev2009, dolaptchiev2013}, 
where the same leading-order equations are derived. They used a different small parameter
$\epsilon=(a\Omega^2/g)^3$ instead of the Rossby number $U/fL$,
where $a$ is the earth's radius and $\Omega$ is the earth's 
rotation frequency, thus the detailed derivations toward the leading-order equations are different.
On the other hand, \cite{moon2020} derive the same results based on the Rossby number as a
small parameter in the asymptotic analysis following the historical development of theories of quasi-geostrophic
motions in large-scale atmospheric dynamics \citep{pedlosky2013}.
 
The two geostrophic motions coexist in the large-scale atmosphere. 
Hence, the large-scale atmospheric dynamics
should be represented by interactions between the two scales. 
Because these two scales are asymptotically separate, 
we can use a multi-scale 
analysis in spatial and temporal domains. 
Based on the Rossby number in the planetary scale 
$\epsilon=U/(f_0L_D)$, we can introduce the two scales, 
$(X,Y,\tilde{t})$ for the planetary scale and $(x,y,t)$ 
for the quasi-geostrophic (QG) one, 
where $(x,y,t)=\epsilon^{1/2}(X,Y,\tilde{t})$. Thus, the time and
spatial derivatives are scaled as
\begin{align}
 \frac{\partial}{\partial t}
  & \ \rightarrow\ \frac{\partial}{\partial \tilde{t}} + 
    \epsilon^{-1/2}\frac{\partial}{\partial t} \nonumber \\
  \frac{\partial}{\partial x}
  & \ \rightarrow\ \frac{\partial}{\partial X} + 
    \epsilon^{-1/2}\frac{\partial}{\partial x} \nonumber \\
  \frac{\partial}{\partial y}
  & \ \rightarrow\ \frac{\partial}{\partial Y} + 
    \epsilon^{-1/2}\frac{\partial}{\partial y}\, . \nonumber
\end{align}
The two scales are separated by $\epsilon^{1/2}$ which comes from
the estimation that $L/L_D \sim \epsilon^{1/2}$ in the large-scale atmosphere in the earth. 
The internal Rossby deformation radius $L$ is around $1000$km in mid-latitudes and
the external (barotropic) one $L_D$ around $3000$km. 

A regular perturbation analysis of the primitive equations (1-4) \citep{moon2020}
leads to 
\begin{align}
  &u_L=-\frac{\partial P_L}{\partial Y},\quad v_L=\frac{\partial P_L}{\partial
    X},\quad \rho_L=-\frac{1}{\rho_s}\frac{\partial}{\partial z}\left(\rho_s
  P_L\right) \label{eq:pl_b} \\ 
  &u_0=-\frac{\partial P_0}{\partial y},\quad v_0=\frac{\partial
    P_0}{\partial x},\quad \rho_0=-\frac{1}{\rho_s}\frac{\partial}{\partial
    z}\left(\rho_s P_0\right) \label{eq:synop_b} \\ 
    & \frac{\partial w_L}{\partial z} 
  -\frac{1}{H}w_L\ =\ \beta_L v_L \label{eq:sverdrup} \\ 
  & \frac{\partial\Theta_L}{\partial\tilde{t}} +
  u_L\frac{\partial\Theta_L}{\partial X} + v_L\frac{\partial\Theta_L}{\partial
    Y} + w_L\left(\frac{\partial\Theta_L}{\partial z} + S\right)\ =\ \nonumber
  \\ & \ -\left(\frac{\partial}{\partial t} + u_L\frac{\partial}{\partial
    x} + v_L\frac{\partial}{\partial y}\right)\Theta_0 -
 \left(u_0\frac{\partial}{\partial X}+v_0\frac{\partial}{\partial Y}\right)\Theta_L \nonumber \\
  &-\frac{\partial}{\partial x}\Big(u_0\Theta_0\Big) -
  \frac{\partial}{\partial y}\Big(v_0\Theta_0\Big) 
  -w_1\left(\frac{\partial\Theta_L}{\partial z} + S\right) + Q_L \label{eq:multi_hq}\\ 
  & \frac{\partial}{\partial t}\nabla^2P_0 + (u_L+u_0)\frac{\partial}{\partial
    x}\nabla^2P_0 + (v_L+v_0)\frac{\partial}{\partial y}\nabla^2P_0 + \beta v_0
  \nonumber \\ & \quad =\left(\frac{\partial}{\partial z} -\frac{1}{H}\right)w_1\ \label{eq:multi_qg},
\end{align}
where $\frac{1}{H}=-\frac{1}{\rho_s}\frac{d\rho_s(z)}{dz}$ and 
$S =\frac{1}{F\Theta_s}\frac{d\Theta_s(z)}{dz}$. The Burger number $F$ is 
defined as $L^2/L^2_D$ where $L$ is an internal Rossby number and 
$L_D$ the external Rossby number. These $L$ and the numbers 
$0$ and $1$ are used as the subscripts to
represent planetary-scale and synoptic-scale variables, respectively. The number $0$ (the number $1$) represents
the leading-order (the next order) in the synoptic-scale.
It is assumed that
planetary-scale (synoptic-scale) variables are only dependent upon 
planetary-scale (synoptic-scale) coordinates, $X$, $Y$, and $\tilde{t}$
($x$, $y$, and $t$). The vertical coordinate $z$ is used for the
both scales.
The equations (\ref{eq:pl_b}-\ref{eq:multi_qg}) show the dynamics of the two scales and their mutual interactions. 
The detailed derivation and discussions are 
found in \cite{moon2020}. The equations (\ref{eq:pl_b}-\ref{eq:sverdrup})
describe that the two scales satisfy the basic balances, geostrophic 
and hydrostatic balances. The 
continuity equation (\ref{eq:sverdrup}) in the planetary scale includes the $O(1)$ beta effect, 
which is known as the Sverdrup 
relation. The two scales contribute to the energy flux 
balance in the heat equation (\ref{eq:multi_hq}) with their own temporal
and spatial scales. The last equation (\ref{eq:multi_qg}) is the vorticity equation 
governing the quasi-geostrophic dynamics. 
Here the planetary geostrophic motion provides a mean field 
for the development of vorticity on the Rossby deformation scale.
\cite{dolaptchiev2013} also considered a similar multi-scale 
analysis to represent the interactions between planetary and synoptic scales, which
leads to equivalent results. However, their main focus lies on vorticity dynamics at the 
two scales, and hence a planetary vorticity equation
is derived from the combination of the heat equation and the Sverdrup relation along with the quasi-geostrophic potential vorticity equation.
Our focus is in the planetary-scale heat equation with thermal forcing $Q_L$. 
Lying between the simple energy flux balance model \citep{north1981} and the quasi-geostrophic dynamics, the planetary heat equation
together with basic dynamic balances connects the planetary thermal heat flux to fluid dynamics on planetary scales.

The above equations become simpler when the planetary-scale 
thermodynamic forcing $Q_L$ is zonally homogeneous.
The planetary scale preserves that symmetry, in which case 
all $X$-derivative terms vanish. Hence, the zonal means of the planetary
meridional and vertical velocities $v_L$ and $w_L$ become zero. 
This yields
\begin{align}
 & u_L\ =\ -\frac{\partial P_L}{\partial Y} \\ &
  \frac{\partial\Theta_L}{\partial\tilde{t}}\ =\ -\left(\frac{\partial}{\partial t} +
  u_L\frac{\partial}{\partial x}\right)\Theta_0 -\frac{\partial}{\partial
    x}\Big(u_0\Theta_0\Big) \nonumber \\ & 
  -\frac{\partial}{\partial y}\Big(v_0\Theta_0\Big) -
  w_1\left(\frac{\partial\Theta_L}{\partial z} 
  + S\right)-v_0\frac{\partial\Theta_L}{\partial Y} + Q_L \label{eq:pla_heat}\\ &
  \frac{\partial}{\partial t}\nabla^2P_0 + (u_L + u_0)\frac{\partial}{\partial
    x}\nabla^2P_0 + v_0\frac{\partial}{\partial y}\nabla^2P_0 + \beta v_0\ =
  \nonumber \\ & \qquad \qquad \left(\frac{\partial}{\partial z} -
  \frac{1}{H}\right)w_1\, \label{eq:vorticity}.
\end{align}
We can introduce a planetary-scale time- and spatial-average to the QG variables under the
assumption that the time- and spatial-average of a QG variable is close to zero. This implies that
the overall effect of synoptic scales on the planetary-scale motions is represented by the 
temporal and spatial average of QG-scale fluxes. 
In particular, it is important to consider a planetary-scale spatial-average over the terms
containing the QG-scale spatial derivative such as $\partial/\partial x$ and $\partial/\partial y$.
Let's consider a large horizontal length $l$ for the spatial-average in the synoptic-scale and then the meridional average  
of $\frac{\partial F}{\partial y}\Big\rvert_{Y_0}$, where $Y_0$ represents a position in the planetary-scale coordinate, is then
interpreted as
\begin{align}
\frac{\partial F}{\partial y}\Big\rvert_Y &= \frac{1}{2l}\int_{Y_0-l}^{Y_0+l}\frac{\partial}{\partial y}F dy \nonumber \\
 &\simeq \lim_{\Delta Y\to 0} \frac{\epsilon^{1/2}}{2\Delta Y}
 \left(F(Y_0+\Delta Y)-F(Y_0-\Delta Y)\right) \nonumber \\
 & = \epsilon^{\frac{1}{2} }\frac{\partial}{\partial Y} F.
\end{align}
Here $l$ and $\Delta Y$ are non-dimensional lengths in synoptic- and planetary-scale, respectively. 
The same length is expressed in two length scales, i.e., $l^* = \Delta Y^*$ where $*$ is used to represent 
dimensional quantities, which is same as $Ll = L_D\Delta Y$. Thus, $l = \Delta Y/\epsilon^{1/2}$ where
$L/L_D =\epsilon^{1/2}$ is used. Due to the asymptotic difference between the two scales, the length $l$
in the synoptic-scale is approximated by the limit of the planetary-scale length $\Delta Y$ toward zero. 

The heat equation (\ref{eq:pla_heat}) and the QG vorticity equation (\ref{eq:vorticity}), 
after taking the planetary-scale time-average $\overline{(\,\cdot\,)}$, become
\begin{align}
&\frac{\partial\Theta_L}{\partial\tilde{t}} = Q_L-\frac{\partial}{\partial x}\overline{u_0\Theta_0}
-\frac{\partial}{\partial y}\overline{v_0\Theta_0}-\overline{w_1}\left(\frac{\partial\Theta_L}{\partial z}+S\right) \label{eq:tmean_theta} \\
&\frac{1}{\rho_s}\frac{\partial}{\partial z}(\rho_s\overline{w_1})
=\frac{\partial}{\partial x}\overline{u_0\nabla^2P_0}+\frac{\partial}{\partial y}\overline{v_0\nabla^2P_0} \label{eq:tmean_vorticity},
\end{align}
where the time average of linear terms in synoptic scales such as $\partial\Theta_0/\partial t$ and $\partial\Theta_0/\partial x$
is assumed to be zero and the nonlinear terms representing heat and momentum fluxes are considered as non-zero average terms.
Using $\nabla^2P_0 = \partial v_0/\partial x -  \partial u_0/\partial y$, we find that the equation (\ref{eq:tmean_vorticity}) is equivalent to 
\begin{align}
\frac{1}{\rho_s}\frac{\partial}{\partial z}(\rho_s\overline{w_1})
=\left(\frac{\partial^2}{\partial x^2}-\frac{\partial^2}{\partial y^2}\right)\overline{u_0v_0}
-\frac{\partial^2}{\partial x \partial y}\overline{u^2_0-v^2_0}.
\end{align} 
  
Now, we can take the planetary-scale spatial average over the equations (\ref{eq:tmean_theta}, \ref{eq:tmean_vorticity}), 
and combine the two by replacing $\overline{w_1}$ by the horizontal vorticity convergence, 
which leads
to
\begin{align} \label{eq:full_planetary_heat}
  & \frac{\partial\Theta_L}{\partial\tilde{t}}\ =\ Q_L 
  - \epsilon^{1/2}\left(\frac{\partial}{\partial X}
  \overline{u_0\Theta_0} 
  +\frac{\partial}{\partial Y}
  \overline{v_0\Theta_0} \right) 
  \nonumber \\ & 
  +\epsilon\left(\frac{\partial\Theta_L}{\partial z} + S\right)
  \left(\frac{\partial^2}{\partial X^2}-\frac{\partial^2}{\partial Y^2}\right)\left(\frac{1}{\rho_s}\int_{0}^{z}\!\rho_s
  \overline{u_0v_0}\,{\rm d}z'\right) \\ \nonumber & 
  -\epsilon\left(\frac{\partial\Theta_L}{\partial z} + S\right)
   \frac{\partial^2}{\partial X \partial Y}\left(\frac{1}{\rho_s}\int_{0}^{z}\!\rho_s 
   (\overline{u^2_0}-\overline{v^2_0})\,{\rm d}z'\right)\,.
\end{align}

The equation (\ref{eq:full_planetary_heat}), after taking the zonal-average $\langle\,(\,\cdot\,)\,\rangle$, becomes
\begin{align} \label{eq:sym_heat}
 & \frac{\partial\Theta_L}{\partial\tilde{t}}\ =\ Q_L - \epsilon^{1/2}\frac{\partial}{\partial Y}
  \Big\langle\overline{v_0\Theta_0}\Big\rangle \nonumber \\ & \qquad \qquad
  -\epsilon\left(\frac{\partial\Theta_L}{\partial z} + S\right)
  \frac{\partial^2}{\partial Y^2}\left(\frac{1}{\rho_s}\int_{0}^{z}\!\rho_s
  \Big\langle\overline{u_0v_0}\Big\rangle\,{\rm d}z'\right)\, ,
\end{align}
where the dominant balance after ignoring the momentum flux contribution in the $O(\epsilon)$ order is
\begin{align} \label{eq:planetary_heat}
  \frac{\partial\Theta_L}{\partial\tilde{t}}\ \simeq \ Q_L - \epsilon^{1/2}\frac{\partial}{\partial Y}
  \Big\langle\overline{v_0\Theta_0}\Big\rangle.
\end{align}

In the equation (\ref{eq:planetary_heat}), we can consider an asymptotic solution 
$\Theta_L \simeq [\Theta_L]+\epsilon^{1/2}\eta$, in which case the radiative process $Q_L$ is 
represented approximately as
$Q_L(\tilde{t},\Theta_L,Y) \simeq Q_L(\tilde{t},[\Theta_L],Y)+\epsilon^{1/2}a(Y,\tilde{t})\eta$,
where $a(Y,\tilde{t}) = \partial Q_L(\tilde{t},\Theta_L,Y)/\partial \Theta_L |_{\Theta_L=[\Theta_L]}$.
The leading-order balance is
\begin{align}
\frac{\partial [\Theta_L]}{\partial\tilde{t}} \simeq  [Q_L].
\end{align}
This is understood as a local radiative energy flux balance, and $[\Theta _L]$ represents
the seasonal mean of the potential temperature mainly determined by the seasonally-varying
radiative flux balance $[Q_L]$. The simplest representation of $[Q_L]$ is 
$S_0(1-\alpha)-\sigma [\Theta_L]^4$, 
where $S_0$ is the shortwave radiative flux, 
$\alpha$ is the local albedo, and $\sigma [\Theta_L]^4$ is 
the outgoing longwave radiative flux, but we have to consider other heat
fluxes such as incoming longwave radiative flux and turbulent sensible and 
latent heat flux. In the planetary-scale heat equation, the leading-order governing physics
is the local energy flux balance contributing mainly to a seasonal cycle of 
the potential temperature. 

The $O(\epsilon^{1/2})$ order balance is
\begin{align} \label{eq:eta}
 \frac{\partial\eta}{\partial\tilde{t}} = a(Y,\tilde{t})\eta - \frac{\partial}{\partial Y}
\Big\langle\overline{v_0\Theta_0}\Big\rangle.
\end{align}
The fluctuation $\eta$ around the seasonal cycle $[\Theta_L]$ is controlled by the seasonal
sensitivity of the radiative processes $a(Y,\tilde{t})$ and the meridional heat flux convergence
induced by synoptic eddies. \cite{moon2017} focus on the monthly-average variability 
of surface air temperature 
based on a periodic non-autonomous stochastic 
differential equation $\dot{\eta}=a(\tilde{t})\eta+N(\tilde{t})\xi(\tilde{t})$, 
where $a(\tilde{t})$ is equivalent to the $a(Y,\tilde{t})$ in (\ref{eq:eta}), 
$\xi(\tilde{t})$ a white noise mimicking the overall
effect of weather-related processes and $N(\tilde{t})$ is a monthly-varying 
amplitude of noise. The noise forcing $N(\tilde{t})\xi(\tilde{t})$ 
can be understood as an approximation of the residual forcing $R(Y,\tilde{t})$
which can be considered as a contribution of short-time processes in
the equation (\ref{eq:eta}).
The monthly statistics including variance and lagged correlations 
in a given monthly-averaged data such as surface sea temperature 
or tropical climate indexes 
is regenerated by the periodic non-autonomous stochastic model with an appropriate
choice of the two periodic functions $a(\tilde{t})$ and $N(\tilde{t})$. 
In particular, the positive sign of the $a(\tilde{t})$ 
implies existence of positive feedbacks which 
magnifies the magnitude of a given perturbation. 
For example, when the stochastic model 
is applied to the Nino 3.4 index, the $a(\tilde{t})$ 
is positive from July to November,
which shows the seasonality of the Bjerknes feedback. 
The construction of the monthly sensitivity 
$a(\tilde{t})$ enables us to detect how a positive feedback shapes 
the monthly statistics of a climate variable. It is essential to understand 
how phase locking and seasonal predictability barrier of climate phenomena are associated
with the magnitude and timing of positive feedback.

The equation (\ref{eq:eta}) is an extension of the one-dimensional stochastic model 
considering meridional variation. Especially, it represents a strong influence from synoptic-scale
eddies on the planetary-scale temperature.
It is in contrast with \cite{boljka2018a} which provide a similar budget equation
on the planetary scale based on the multi-scale analysis suggested by \cite{dolaptchiev2013}. In their budget equations, the planetary scale
is independent from the synoptic scale, thus the two scales interact indirectly through source terms. On the other hand, the equation (\ref{eq:eta})
tells that the variability of planetary-scale temperature is mainly determined by the turbulent heat flux induced by synoptic eddies. 
One remaining step to close the planetary-scale equations 
is to parameterize 
the meridional heat flux $\Big\langle\overline{v_0\Theta_0}\Big\rangle$
using the planetary-scale temperature $\Theta_L$. 

\section{Emergence of a generalized Langevin equation}

\subsection{Fickian diffusion model}
The meridional heat flux at mid-latitudes under the growth of synoptic waves 
is the major consequence of baroclinic instability.
The baroclinic instability is initiated from an unstable mean state measured 
by the vertical shear of zonal mean wind $\partial u_L/\partial z$.
The growth of baroclinic waves by a baroclinic instability starts from near surface, 
which induces a meridional heat flux. Hence, the meridional gradient 
of the planetary temperature $\partial\Theta_L/\partial Y$ is 
strongly related to the meridional heat flux
$\Big\langle\overline{v_0\Theta_0}\Big\rangle$.

The simplest representation of the mutual relationship 
between $\partial\Theta_L/\partial Y$ and 
$\langle\overline{v_0\Theta_0}\rangle$
is a turbulent flux gradient parameterization, 
$\langle\overline{v_0\Theta_0}\rangle 
\simeq -K\partial\Theta_L/\partial Y$, where $K$
is a constant \citep{north1981}. If we use this parameterization 
in the equation (\ref{eq:planetary_heat}), 
this leads to
\begin{align}\label{eq:mean_temp}
  \frac{\partial \Theta_L}{\partial\tilde{t}} 
  \simeq  \epsilon^{1/2}K\frac{\partial^2  \Theta_L}{\partial Y^2}+ Q_L. 
\end{align}
This simple energy flux balance model was first introduced by \cite{sellers1969} and \cite{north1975} 
to include the effect of large-scale atmospheric dynamics 
upon the local energy flux balance. 
The consequence of complicated large-scale atmospheric dynamics especially 
in mid-latitudes is to transfer the surplus of 
energy in low latitudes to high latitudes, 
which is simply approximated by a turbulent meridional diffusion. 

The temporal 
and spatial evolution of the perturbation $\eta$ is represented by
\begin{align}\label{eq:main_eta}
 \frac{\partial \eta}{\partial \tilde{t}} = K\frac{\partial^2 \eta}{\partial Y^2}+a(Y,\tilde{t})\eta+R(Y,\tilde{t}),
\end{align} 
where we include the $R(Y,\tilde{t})$ for the contribution of short-time processes.  
We can consider two boundary conditions in meridional coordinate, 
$\frac{\partial\eta}{\partial Y}(\cdot,Y=0) = \frac{\partial\eta}{\partial Y}(\cdot,Y=1) = 0$
implying there is no meridional heat flux near the pole and the tropical areas. 

For the $Y$-direction diffusion operator, consider the eigenvalue problem,
\begin{align}\label{eq:eigenq}
 K\frac{\partial^2}{\partial Y^2} H_n = -\lambda_n H_n
\end{align}
with $\frac{\partial H_n}{\partial Y}(Y=0) = \frac{\partial H_n}{\partial Y}(Y=1) = 0$. 
Here, $H_n=A_n\text{cos} (n \pi Y)$ with $\lambda_n=Kn^2\pi^2$ and $n=0,1,2,\cdots$. 
Hence, the temporal and spatial perturbation 
of the potential temperature anomaly $\eta$ can be represented 
by an infinite series of the eigenfunctions with time-varying coefficients,
\begin{align}
 \eta = \sum_{n=0}^{\infty} x_n(\tilde{t}) H_n(Y).
\end{align}
 Consider first the case that $a(Y,\tilde{t})$ is a constant $-\gamma$ and that
 the sum of short-time processes $R(Y,\tilde{t})$ is represented 
 as a sum of the same eigenfunctions, $R = \sum_{n=0}^{\infty}R_n(\tilde{t})H_n(Y)$.
 
 The time-varying coefficient $x_n(\tilde{t})$ then satisfies 
 \begin{align} \label{eq:langevin}
 \frac{dx_n(\tilde{t})}{d\tilde{t}} = -\left(\lambda_n+\gamma\right)x_n(\tilde{t})+R_n(\tilde{t}).
 \end{align}
 If we consider the last term as a stochastic noise, in particular, 
 a Gaussian white noise, this becomes a Langevin equation 
 with a decay time scale $1/(\lambda_n+\gamma)$. 
 The time scale becomes shorter as $n$ increases, which indicates that 
 the first several modes representing large-scale motions 
 dominate in the overall fluctuations.
 When $n$ is equal to $0$, the Langevin equation 
 becomes $\dot{x_0} = -\gamma x_0 + R_0(\tilde{t})$, where $\gamma$ represents
 the seasonal sensitivity of local energy flux balance mainly 
 dominated by radiative-convective equilibrium in land and ocean heat
 flux in ocean boundary layer. The formalism used in this argument 
 was first introduced by \cite{manucharyan2016b} to explain the 
 internal variability caused by interactions between 
 meso-scale eddies and mean fields 
 under the framework of the Gent-McWilliam parameterization.
 
 The simple Langevin equation was introduced to climate 
 science by \cite{hasselmann1976} 
 to interpret climate variability in terms of stochastic 
 dynamics. The deterministic part is understood 
 as a stabilizing process around a mean state 
 and the stochastic noise as the 
 effect of weather. Climate is understood as a combination of 
 the stability of slowly-evolving backgrounds and short-time 
 processes approximated as a noise. The interpretation was useful 
 to rationalize the ubiquitous emergence of red noise spectra from
 climate variables including sea surface 
 temperature (SST) \citep{frankignoul1977}. In a symmetric hemisphere, 
 the meridional variation of mean near-surface temperature 
 could be constructed by the equation (\ref{eq:mean_temp}) 
 and the fluctuation around the mean be represented 
 by the equation (\ref{eq:langevin}), which might be consistent 
 with the usage of an AR1 process explaining the 
 zonal index variability \citep{lorenz2001}. 
 
 \subsection{Quasi-oscillatory behavior of the BAM}
 \cite{thompson2014} (TB) studied quasi-periodic variability 
 of the Southern Hemisphere baroclinic annular mode (BAM).
 Instead of the red noise spectrum originated from a simple linear Langevin equation, 
 the BAM shows a clear sign of quasi-oscillation 
 in its power spectrum. This indicates that the simple Langevin equation 
 leading to a red-noise spectrum is not adequate to describe the quasi-periodic 
 fluctuation.
 
 To explain a periodic behavior of meridional temperature gradient 
 and poleward heat flux,  
 TB introduce mutual interaction and feedback 
 between the meridional temperature gradient $b \equiv \partial\eta/\partial Y$ 
 and the poleward heat flux $\langle \overline{v_0\Theta_0}\rangle$. 
 Baroclinic instability theory tells us that 
 the growth rate of the baroclinic waves 
 is proportional to the meridional temperature gradient \citep{lindzen1980}, which 
 could be represented by
 \begin{align} \label{eq:vt}
  \frac{d}{dt}\langle\overline{v_0\Theta_0}\rangle = -\alpha\langle b\rangle +\epsilon(t),
 \end{align}   
 where $\alpha$ is a constant representing the amplitude of the feedback between
 the baroclinicity and the meridional heat flux \citep{hoskins1990}.
 Here, the noise forcing $\epsilon(t)$ is added to include 
 the effect of chaotic weather. 
 At the same time, TB consider a feedback of 
 the poleward heat flux on the meridional temperature gradient, 
 which simply assumes that the meridional temperature gradient $\langle b\rangle$
 increases linearly with respect to the poleward heat flux 
 $\langle \overline{v_0\Theta_0} \rangle$. 
 Thus, the second equation in TB is
 \begin{align}\label{eq:b}
  \frac{db}{dt} = \mu\langle\overline{v_0\Theta_0}\rangle - \frac{b}{r},
 \end{align}
 where $\mu$ is the degree of the feedback and $r$ is 
 a recovering time-scale of the meridional temperature gradient due to 
 diabatic processes and vertical motions. 
 The combination of (\ref{eq:vt}-\ref{eq:b}) 
 could generate an oscillation whose period is 
 determined by the relevant coefficients. 
 While ignoring the noise forcing in the equation (\ref{eq:vt}), 
 combining the two equations  \citep{thompson2014}
 leads to
 \begin{align} \label{eq:osi_b}
  \frac{d^2b}{dt^2} + \frac{1}{r}\frac{db}{dt} +\alpha\mu b = G,
 \end{align} 
 where the $G$ is the time-derivative of the short-time scale forcing $\epsilon(t)$. 
 Oscillatory behavior comes out when 
 $\frac{1}{4r^2} - \alpha\mu < 0$. TB used $r \simeq 4$ days 
 and $\alpha\mu \simeq 0.05 \text{days}^{-2}$ 
 to generate a dominant peak of the power spectrum of 
 poleward heat flux around 20 days. 
 They suggest a mutual interaction between 
 $\langle\overline{v_0\Theta_0}\rangle$ and $\partial\eta/\partial Y$ 
 with the damping represented by a Newtonian cooling. 
 In the model, it is hard to guess what determines 
 the damping time-scale $r$ and the feedback parameter $\alpha\mu$.
 
 By a simple comparison, the equation (\ref{eq:b}) 
 is equivalent to the anomalous heat equation (\ref{eq:eta}) and then
 the equation (\ref{eq:vt}) provides a relationship 
 between the poleward heat flux and the meridional temperature gradient. 
 Even though the equation (\ref{eq:b}) is likely to be derived from the anomalous heat
 equation, the damping term  $-b/r$ is not clearly understood physically.
 To understand the relationship between $b$ and $\langle\overline{v_0\Theta_0}\rangle$
 on intra-seasonal time-scales, it seems necessary what determines the damping time
 scale $r$. 
 
 We can incorporate the equation (\ref{eq:vt}) 
 into the equation (\ref{eq:eta}) 
 after taking time-derivative on the equation (\ref{eq:eta}).
 This results in
 \begin{align} \label{eq:eta_diabatic}
 \frac{\partial^2\eta}{\partial\tilde{t}^2} = \alpha\frac{\partial^2\eta}{\partial Y^2}-\gamma\frac{\partial\eta}{\partial\tilde{t}}+\frac{\partial R}{\partial\tilde{t}},
 \end{align}
 where $-\gamma$ is used instead of the $a(Y,\tilde{t})$ for simplicity. 
 The constant $\gamma$ implies the damping time scale of seasonal energy 
 flux balance. In land, the outgoing longwave radiative flux dominantly 
 controls the time scale.
 Similarly, we can consider the eigenvalue problem for the diffusion operator,
 \begin{align}
  \alpha\frac{\partial^2}{\partial Y^2} H_n = -\lambda_n H_n,
 \end{align}
 where $\eta =  \sum_{n=0}^{\infty} x_n(\tilde{t}) H_n(Y)$. 
 We obtain the time-evolution equation for $x_n(\tilde{t})$,
 \begin{align} \label{eq:rad_osi_xn}
 \frac{d^2x_n}{d\tilde{t}^2}+\gamma\frac{dx_n}{d\tilde{t}}+\lambda_n x_n = \frac{dR_n}{d\tilde{t}}, 
 \end{align}
 where $R= \sum_{n=0}^{\infty} R_n(\tilde{t}) H_n(Y)$ is used. 
 The equation for $x_n$ is similar to the equation (\ref{eq:osi_b}), which 
 suggests that the damping time scale is equivalent to $1/\gamma$. 
 Physically, the $\gamma$ is introduced as a sensitivity 
 of the radiative energy flux balance. In terms of time-scale, 
 it could be interpreted as a time-scale to return to a climatological
 seasonal cycle. \cite{moon2017} developed a time-series method 
 to construct the $\gamma$ 
 from monthly-averaged surface air temperature. 
 The time scale of the surface energy flux balance in the Southern Hemisphere 
 is around 1.5 months, which is much larger than 4 days.  
 Therefore, the damping time scale $r$ does not come from diabatic 
 processes related to radiative fluxes.
 
 \subsection{Eddy memory and generalized Langevin equation}
 
 The damping time scale introduced to explain the oscillatory behavior 
 of the baroclinic mode in Southern Hemisphere is much shorter than
 that of the mean seasonal cycle of radiative processes. 
 It is plausible that the time scale is related to synoptic eddies, 
 rather than any external
 influences which have longer time-scales. 
 Synoptic eddies generated from the baroclinic instability 
 due to an unstable background undergoes a specific energy cycle 
 with a zonal mean steady state. The baroclinic instability enables the synoptic 
 eddies to extract energy from the zonal mean state, 
 from which the poleward heat flux increases near surface. 
 The growing waves propagate upward
 and equatorward and meet critical latitudes where phase speeds 
 of waves are same as mean wind. 
 The waves break and give their energy 
 back to the mean state by momentum flux. 
 This baroclinic wave life cycle is complete in a few days. 
 
 The overall effect of the baroclinic wave life cycle on the meridional heat flux 
 is represented by the planetary potential temperature anomaly $\eta$ as
 $ \langle\overline{v_0\Theta_0}\rangle
 = -K\frac{\partial \eta}{\partial Y}$. This Fickian diffusion approximation is well
 applied when the time evolution of the planetary variable is 
 much slower than the turn-over time-scale of synoptic eddies. 
 On intraseasonal time scales, however, 
 the time scale for the evolution of planetary-scale variables 
 is not clearly separated from that for the baroclinic wave life cycle. The advective timescale for the synoptic waves
 is around 1.2 days (where we use $L=1000$km and $U=10$m/s) and the same timescale for the planetary waves
 around 3.5 days (where we use $L_D=3000$km and $U=10$m/s).  A complete cycle of baroclinic wave lifecycle takes
 around 3-4 days, hence the timescale for the planetary-scale motion is comparable with that for the baroclinic wave life cycles.
 It is questionable to apply the 
 Fickian diffusion as a parameterization 
 of the poleward  heat flux in these time scales.
 
 Non-Fickian approximation
 for turbulent transport or diffusion has been 
 a central topic in turbulent closure problems \citep{orszag1970}. 
 For a transient and chaotic turbulence, 
 the Fickian approximation is not enough to capture 
 non-local and non-Gaussian nature of turbulence. 
 One of the simplest approach is called the minimal $\tau$
 approximation, where the third-order momentum represented 
 as a forcing for the time-evolution of second-order moments is approximated as a 
 damping term with the timescale $\tau$ \citep{brandenburg2004, brandenburg2005}. 
 This is equivalent to apply an integral kernel instead of 
 a constant diffusivity with a finite memory \citep{hubbard2009}.
 
 The main idea is applied to the time-evolution of the poleward heat flux of synoptic-scale waves, i.e.,
 \begin{align}
     \frac{\partial}{\partial t}\langle v_0\Theta_0 \rangle = 
     \langle \frac{\partial v_0}{\partial t}\Theta_0\rangle
     + \langle v_0\frac{\partial \Theta_0}{\partial t}\rangle.
 \end{align}
 The time-evolution equations for the synoptic-scale meridional velocity $v_0$ and potential temperature $\Theta_0$ \citep{moon2020} are
 \begin{align}
     &\frac{\partial v_0}{\partial t}=-(u_L+u_0)\frac{\partial v_0}{\partial x}
     -v_0\frac{\partial v_0}{\partial y}
     -u_1-\frac{\beta y}{\epsilon^{1/2}}u_0
     -\frac{\partial P_1}{\partial x} \nonumber \\
     &\frac{\partial\Theta_0}{\partial t}=-\frac{\partial\eta}{\partial\tilde{t}}
     -(u_L+u_0)\frac{\partial\Theta_0}{\partial x}-v_0\frac{\partial\eta}{\partial Y} \nonumber \\
     &-v_0\frac{\partial\Theta_0}{\partial y}-w_1\left(\frac{\partial\Theta_L}{\partial z}+S\right).
 \end{align}
 Hence, 
 \begin{align}
     \frac{\partial}{\partial t}\langle v_0\Theta_0\rangle
     =-\langle v^2_0\rangle\frac{\partial\eta}{\partial Y} + D,
 \end{align}
 where $D$ contains the terms representing the contribution from synoptic eddies.
 After taking synoptic-time average on both sides, we obtain
 \begin{align}
     \frac{\partial}{\partial\tilde{t}}\langle\overline{v_0\Theta_0}\rangle
     =-\langle\overline{v^2_0}\rangle\frac{\partial\eta}{\partial Y} + \overline{D}.
 \end{align}
 The main assumption of the minimal $\tau$ approximation is that the contribution of the
 synoptic eddies $\overline{D}$ is approximated by a damping of the 
 $\langle\overline{v_0\Theta_0}\rangle$ with the designated time-scale $r$. Therefore,
 \begin{align} \label{eq:minimal_tau}
     \frac{\partial}{\partial\tilde{t}}\langle\overline{v_0\Theta_0}\rangle
     =-K\frac{\partial\eta}{\partial Y}
     -\frac{\langle\overline{v_0\Theta_0}\rangle}{r},
 \end{align}
 where $K \equiv \langle\overline{v^2_0}\rangle$ suggesting how the eddy diffusivity $K$ 
 is related to the second-order statistics of synoptic eddies.
 The minimal $\tau$ approximation for the
 $\langle\overline{v_0\Theta_0}\rangle$ seems to be equivalent to the equation (\ref{eq:vt})
 in the TB. The feedback parameter $\alpha$ could be understood as 
 $K$. TB suggests the relationship based on the result
 from the baroclinic instability. On the other hand, we derived a similar one by the 
 simplest non-Fickian approximation as a closure of turbulent eddies. 
 TB included a random forcing
 in the relationship to include unresolved processes, 
 but the randomness coming from short-time
 processes is considered in the heat equation in our derivation. 
 The above parameterization can be 
 represented by an integral form, i.e.,
 \begin{align} \label{eq:minimal}
     \langle\overline{v_0\Theta_0}\rangle = -K\frac{\partial}{\partial Y}
     \int_{-\infty}^{\tilde{t}}\eta\, \text{exp}\left(-\frac{\tilde{t}-\tilde{t}'}{r}\right)d\tilde{t}'.
 \end{align}
 The integral form originates from integrating (\ref{eq:minimal_tau})
 for the $ \langle\overline{v_0\Theta_0}\rangle$ with respect to the time $\tilde{t}$. 
 From this integral form, we can see that the poleward heat flux is the result of accumulating the
 baroclinicity during a certain time until present. It represents the memory effect 
 caused by synoptic eddies.
 
 The minimal $\tau$ approximation was tested in direct simulations 
 of isotropic 3D turbulence \citep{brandenburg2004}. The statistics of a passive scalar
 are compared between direct simulations and the parameterized equations,
 where decent matches are obtained. In particular, the parameterization transforms 
 the main parabolic tracer equation to a wave equation leading to an oscillatory behavior.
 The simulations show decayed oscillation 
 with a certain choice of the damping time-scale $r$.
 In an idealized Beaufort Gyre numerical simulation, \cite{manucharyan2017} introduced 
 a finite memory kernel instead of a constant diffusivity 
 in the Gent-McWilliam parameterization, which generates the quasi-periodic variability 
 in the eddy-mean interaction. The minimal $\tau$ approximation can be understood as 
 a finite memory effect represented by an integral kernel.

 Taking account of memory effect of eddies, we can introduce 
 an integral kernel $\kappa(t-t')$ on the parameterization 
 of the meridional heat flux,
 \begin{align}
 \langle\overline{v_0\Theta_0}\rangle
 &=-K\frac{\partial}{\partial Y}\int_{-\infty}^{t}\eta(t')\kappa(t-t')dt', \nonumber \\
  &= -K\frac{\partial \eta^*}{\partial Y} 
 \end{align}
 which is a generalization of the minimal $\tau$ approximation in the equation (\ref{eq:minimal})
 where $\eta^*$ is an effective temperature anomaly defined by
 \begin{align}
 \eta^* \equiv \int_{-\infty}^{t}\eta(t')\kappa(t-t')dt'.
 \end{align} 
 
 Thus, the equation (\ref{eq:main_eta}) becomes
 \begin{align}
 \frac{\partial\eta}{\partial\tilde{t}} 
 = K\frac{\partial^2\eta^*}{\partial Y^2}-\gamma\eta+R(Y,\eta),
 \end{align}
 where we assume that $a(Y,\tilde{t})=-\gamma$ for simplicity. 
 Furthermore, if we expand the $\eta^*$ using the eigenfunctions 
 of the diffusion operator,
 $\eta^* = \sum_{n=0}^{\infty}x^*_n(\tilde{t})Q_n(Y)$, and the residual forcing, $R= \sum_{n=0}^{\infty} R_n(\tilde{t}) Q_n(Y)$,
 each time-dependent coefficient $x_n$ satisfies 
 \begin{align} \label{eq:gle_xn}
 \frac{dx_n}{d\tilde{t}} &= -\lambda_n x^*_n-\gamma x_n+R_n \nonumber \\
    &=-\lambda_n \int_{-\infty}^{\tilde{t}}x_n(\tilde{t}')\kappa(\tilde{t}-\tilde{t}')d\tilde{t}-\gamma x_n+R_n,
 \end{align}
 which is a generalized Langevin equation \citep{zwanzig1973}. 
 The generalized Langevin equation contains memory terms representing 
 the effect of past states, which is generally originating from
 interactions with short-time scale components. 
 
 The complicated chaotic processes to generate memory are captured 
 by the memory kernel $\kappa(\tilde{t}-\tilde{t}')$. The dynamics represented by
 a generalized Langevin equation is dependent upon the choice of integral kernel. 
 We can follow the theme of the $\tau$ approximation, using an integral kernel 
 with a finite-time  memory. It has been suggested that 
 baroclinic eddies generated by baroclinic instability are eventually organized 
 to maintain a mid-latitude jet characterized by meridional temperature gradient \citep{lorenz2001,robinson2000}. 
 This implies that the effect of baroclinic eddies upon the meridional heat flux 
 is limited within a finite time-scale, which leads to as an approximation
 \begin{align}
 \kappa(\tilde{t}-\tilde{t}') = \text{exp}\left(-\frac{\tilde{t}-\tilde{t}'}{r}\right).
 \end{align}
 Here $r$ represent the finite memory of the baroclinic eddies, 
 which is equivalent to $dx^*_n/d\tilde{t} = -x^*_n/r+x_n/r$. Based on the specific memory 
 kernel, the equation (\ref{eq:gle_xn}) becomes
 \begin{align}
 \frac{d^2x_n}{d\tilde{t}^2}+\left(\frac{1}{r}+\gamma\right)\frac{dx_n}{d\tilde{t}}
 +\frac{\gamma+\lambda_n}{r}x_n = \frac{R_n}{r}+\frac{dR_n}{d\tilde{t}}.
 \end{align}
 There are two-damping time-scales in the above equation 
 defined by $r$ and $1/\gamma$. The $1/\gamma$ comes from the seasonal heat flux
 balance such that it is approximately $1.5$ months in Southern Hemisphere. 
 Thus, it follows $1/r \gg \gamma$. The eigenvalues $\lambda_n$ come again
 from (\ref{eq:eigenq})   
with $Q'_n(0)=Q'_n(1)=0$ implying no heat flux at both ends, 
which gives us $\lambda_n=Kn^2\pi^2$ and $Q_n(Y)=cos(n\pi Y)$. 
The baroclinic annular mode (BAM) is defined by a dominant mode 
of the meridional heat flux maximized around the center of mid-latitudes. 
This is similar to the mode with $n=1$, $Q_1(Y)$, whose
 derivative $Q'_1(Y)$, has its maximum at the center. 
 The time-dependent coefficient of the mode, $x_1$, satisfies
 \begin{align}\label{eq:sto_osc}
  \frac{d^2x_1}{d\tilde{t}^2}+\frac{1}{r}\frac{dx_1}{d\tilde{t}}+\frac{\gamma+K\pi^2}{r}x_1 = \frac{R_1}{r}+\frac{dR_1}{d\tilde{t}},
 \end{align} 
 which should be understood as an equivalent form to the equation (\ref{eq:osi_b}). 
 The damping time-scale $r=4$ days comes from the memory effect of synoptic eddies
 on meridional heat flux, which seems to be strongly related to 
 the baroclinic wave life cycle.  
 Moreover, $\alpha\mu$ in the equation (\ref{eq:osi_b}) is same as
 $(\gamma+K\pi^2)/r$, which leads to $K \simeq 1/5\pi^2$.

 \begin{figure}[ht]
\centering
\includegraphics[angle=0,scale=0.13,trim= 0mm 0mm 0mm 0mm, clip]{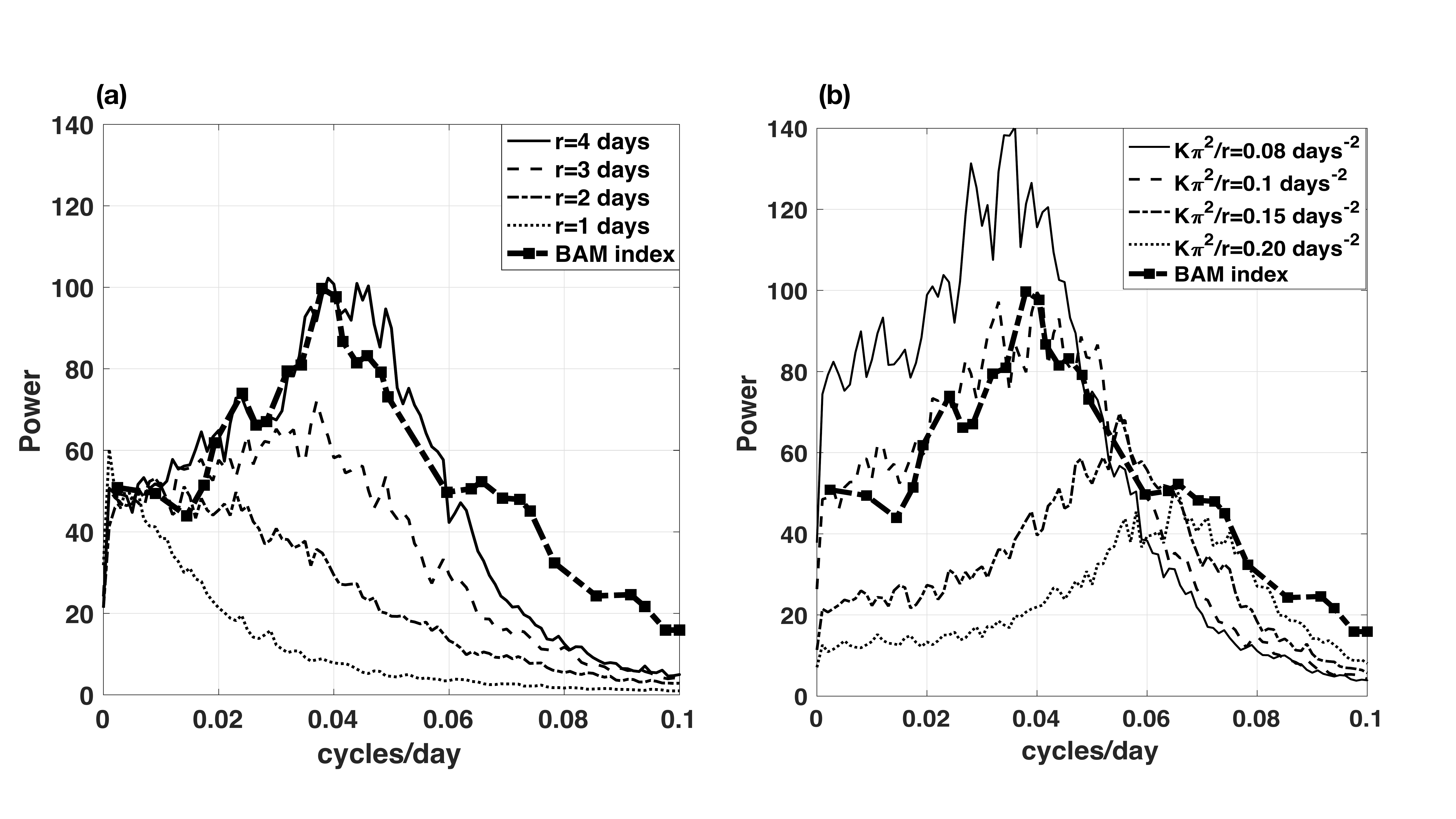}
\caption{Power spectra of the stochastic oscillators represented 
by the equation (\ref{eq:sto_osc}) with several choices 
of the parameters $r$ and $K$ 
along with the power spectrum of the BAM index \citep{thompson2014a}. 
With the fixed
$K=1/5\pi^2$, four different $r$'s are used 
for stochastic realization (a). Similarly,
four different $K$'s are chosen for generating 
power spectra with a fixed $r=4$ (b).}
\label{fig:sto}
\end{figure}
Depending on the choice of the parameters $r$ and $K$, the equation (\ref{eq:sto_osc})
can show decayed oscillatory behavior or exponential decay. Under the assumption that
$R_1/r+dR_1/d\tilde{t}$ is approximately a Gaussian white noise, we can simulate the equation
to generate a stochastic realization. Figure \ref{fig:sto} shows several power spectra 
of stochastic realizations from the equation (\ref{eq:sto_osc}) with different parameters
$r$ and $K$ along with the power spectrum of the BAM index \citep{thompson2014a}. 
With a fixed $K$, four different memory time-scales from 1 day to 4 days
are used for stochastic realization. In figure \ref{fig:sto}(a), we see that longer
memory (4 days) leads to oscillatory behavior, which has the similar power spectrum with that of the BAM index, 
and shorter memory (1 day) to a red noise
spectrum (Fig. \ref{fig:sto}a). Similarly, we fix $r=4$ days and vary $K$, which also shows the characteristics
from stochastic oscillation to red noise process (Fig.\ref{fig:sto}b). Similarly, when $K\pi^2/r = 1\, \text{days}^{-2}$, 
the power spectrum is close to that of the BAM index.
 
The baroclinic wave life cycle results in recovering 
 the vertical shear of the mean jet. This could be an origin of the finite memory effect 
 in meridional heat transfer in planetary-scale, which leads to oscillatory behavior of zonal 
 mean potential temperature. Following the above formalism, the fluctuation of 
 potential temperature anomaly $\eta$ is represented 
 as $\eta = \sum_{n=0}^{\infty} x_n(\tilde{t}) Q_n(Y)$, 
 hence $\partial\eta/\partial Y = \sum_{n=0}^{\infty}x_n(\tilde{t}) dQ_n/dY$. 
 By the thermal wind balance, 
 $\partial U/\partial z \propto \sum_{n=0}^{\infty}x_n(\tilde{t}) dQ_n/dY$, 
 which implies that the vertical shear of the jet in lower levels could show 
 quasi-oscillatory behavior depending on the time-scale of the eddy memory 
 and relevant eigenvalues 
 defining the decay time-scale of a specific normal mode.

The main focus lies on how to parameterize the meridional heat flux 
induced by synoptic eddies $\langle\overline{v_0\Theta_0}\rangle$ using the planetary-scale potential temperature $\Theta_L$. 
Considering that the time scale for synoptic eddies is not much smaller than that of planetary-scale motions, the meridional
heat flux is not entirely determined by an instantaneous potential temperature gradient; instead it is dependent upon the temporal history
of the potential temperature gradient. Hence, the Fickian diffusion approximation is not appropriate to parameterize the meridional
heat flux induced by synoptic eddies.
The simplest way considering the temporal history of the planetary-scale potential temperature
 is to introduce a time integration of the potential temperature with an appropriate memory kernel, 
 which enables to include the past influence of planetary-scale mean fields on
 the turbulent heat flux.
As a result, fluctuations of potential 
temperature can be 
described approximately by a generalized Langevin equation containing a memory term.
The time-delayed effect
represented as an integral kernel in the generalized 
Langevin equation can generate various types of variability 
ranging from simple red-noise process 
to quasi-oscillations and chaos. Hence, an appropriate choice
of integral kernel is crucial to define the statistical 
characteristics of planetary-scale variability.

\section{Conclusion}
The main goal of the study was to rationalize the quasi-oscillatory variability of the BAM \citep{thompson2014} on intra-seasonal time scales 
using the multi-scale representation of primitive equations of the large-scale atmosphere flow \citep{moon2020}. 
Conceptually, the multi-scale approximation is a combination of the energy flux balance in the heat equation and the potential vorticity conservation. We found that the critical issue under the framework is the specification of an appropriate parameterization of the horizontal heat flux convergence from the synoptic and planetary-scale eddy dynamics.  If the heat fluxes are represented using the simplest diffusive (Fickian) parameterization with a constant eddy diffusivity, the planetary-scale equations simplify to an energy flux balance with a meridional turbulent diffusion, similarly to \cite{north1975}. Other types of eddy parameterizations are also possible, of those most notable is 
the residual mean theory of \citet{andrews1976planetary} with a diffusive parameterization of isopycnal eddy fluxes of potential vorticity. 
While the residual mean formulation gives a different perspective on the mean flow dynamics compared to the direct diffusive closure of the heat fluxes \citep{schneider1974}, 
our study questions their equilibrium-type Fickian assumption for the relation between the eddy fluxes and gradients of the considered tracer fields. Under the equilibrium-type eddy parameterizations, 
the main equation in the planetary-scales atmosphere is the heat equation with the basic balances resulting in a parabolic partial differential equation that does not support natural oscillations. The resulting temporal variability of observables obeys the Langevine equation leading to a red-noise spectrum that does not explain the quasi-oscillatory behavior of the BAM. Thus the equilibrium-type eddy parameterizations are likely not appropriate at the intraseasonal timescales and can only be used for representing the relatively long-term average of the planetary-scale dynamics. 

We find that using a non-equilibrium eddy parameterization with the eddy memory effect represented by the delayed integral in the flux-gradient relation as was proposed for mesoscale ocean eddies \citep{manucharyan2017} can explain the oscillatory behavior of planetary-scale atmospheric flows. The rationalization for the eddy memory effect comes from the notion that the overall effect of synoptic eddies is to maintain the vertical shear of a mid-latitude jet \citep{robinson2000}, implying that the eddy energy cycle starting from baroclinic instability near the surface could be considered as a finite timescale process. Thus, we hypothesized that the mid-latitude eddies have a finite memory and introduce a simple exponentially 
decaying memory kernel into the relation between the poleward eddy heat flux and the meridional potential temperature gradient. 
The inclusion of the finite memory kernel directly converts the heat equation from parabolic to wave equation that allows dispersive planetary-scale waves. As a result, instead of the red-noise model, 
the temporal variability of the anomalous potential temperature near the surface now obeys a stochastic oscillator model, with spectral characteristics that are similar to the observables associated with the BAM. 

The key value of our theoretical study is that it highlights the processes and clarifies the physical interpretation of the crucial parameters that may be governing the spectral characteristics of the BAM. Specifically, our study implies that the period of natural oscillations of the BAM is proportional to the geometric mean between the eddy memory and the diffusive equilibration timescales. The ratio of these timescales controls the existence of a pronounced spectral peak while the variance is directly proportional to the external noise variance in the heat equation. 
Validating our key hypothesis about the exponential memory kernel and the finite eddy memory timescale in the relation between the mean flow and eddy heat fluxes using atmospheric models would be the next crucial step towards validating our theoretical arguments about the BAM. Furthermore, understanding the nature of the eddy memory and how its timescale depends on the mean flow or eddy characteristics would be necessary to understand the climate conditions under which the BAM could exhibit oscillatory behaviour. Finally, our results emphasize the importance of the external noise in driving the variance of the BAM and the necessity to understand if it is dependent on the mean flow or acts as an external/independent source of energy.


\begin{acknowledgements}
W. Moon acknowledges the support of Swedish Research Council Grant No. 638-2013-9243.
G.E.M. acknowledges support from the United States Office of Naval Research award N00014-19-1-2421.
HD acknowledge support by the Netherlands Earth System Science Centre (NESSC), financially supported by the Ministry of Education, 
Culture and Science (OCW), Grant No. 024.002.001. 
\end{acknowledgements}

\bibliographystyle{wileyqj}

\end{document}